\documentclass[journal]{vgtc}                
\ifpdf
  \pdfoutput=1\relax                   
  \usepackage{graphicx}                
  \DeclareGraphicsExtensions{.pdf,.png,.jpg,.jpeg} 
\else
  \usepackage{graphicx}                
  \DeclareGraphicsExtensions{.eps}     
\fi%

\graphicspath{{figures/}{pictures/}{images/}{./}} 

\usepackage{microtype}                 
\PassOptionsToPackage{warn}{textcomp}  
\usepackage{textcomp}                  
\usepackage{mathptmx}                  
\usepackage{times}                     
\usepackage{cite}                      
\usepackage{tabu}                      
\usepackage{booktabs}                  

\usepackage{subfig}
\usepackage{enumitem}
\usepackage[utf8]{inputenc}
\usepackage[T1]{fontenc}
\usepackage{amsmath}
\usepackage{amsfonts}
\usepackage{amssymb}
\usepackage{hyperref}

\newcommand{\codeunit}{block}
\newcommand{\Codeunit}{Block}
\newcommand{\codeunitcontainer}{library}

\newcommand{\sys}{Lodestar}
\newcommand{\advisor}{advisor}
\newcommand{\Advisor}{Advisor}

\onlineid{1004}

\vgtccategory{Research}
\vgtcpapertype{System}

\title{Lodestar: Supporting Independent Learning and Rapid Experimentation Through Data-Driven Analysis Recommendations}

\author{Deepthi Raghunandan, Zhe Cui, Kartik Krishnan, Segen Tirfe, Shenzhi Shi,\\ Tejaswi Darshan Shrestha, Leilani Battle, and Niklas Elmqvist, \textit{Senior Member, IEEE}}
\authorfooter{
\item 
 Deepthi Raghunandan, Department of Computer Science, University of Maryland, E-mail: draghun1@umd.edu.
\item
 Zhe Cui, Google, Inc., E-mail: zhecui@google.com.
\item
 Kartik Krishnan, Department of Computer Science, University of Maryland, E-mail: kkrishn1@umd.edu.
\item
 Segen Trife, Department of Information Science, University of Maryland, E-mail: segent@umd.edu.
\item 
 Shenzhi Shi, Department of Computer Science, University of Maryland,  E-mail: sshi1234@umd.edu. 
\item 
 Tejaswi Darshan Shrestha, Department of Computer Science, University of Maryland, -mail: shrestha.tj@gmail.com.
\item 
 Leilani Battle, Department of Computer Science and Engineering, University of Washington, E-mail: leibatt@cs.washington.edu.
\item
 Niklas Elmqvist, College of Information Studies, University of Maryland, E-mail: elm@umd.edu
}

\shortauthortitle{Raghunandan \MakeLowercase{\textit{et al.}}: Lodestar}

\abstract{
    Keeping abreast of current trends, technologies, and best practices in visualization and data analysis is becoming increasingly difficult, especially for fledgling data scientists.
    In this paper, we propose Lodestar, an interactive computational notebook that allows users to quickly explore and construct new data science workflows by selecting from a list of automated analysis recommendations.
    We derive our recommendations from directed graphs of known analysis states, with two input sources: one manually curated from online data science tutorials, and another extracted through semi-automatic analysis of a corpus of over 6,000 Jupyter notebooks.
    We evaluate Lodestar in a formative study guiding our next set of improvements to the tool. 
    Our results suggest that users find Lodestar useful for rapidly creating data science workflows.
} 


\CCScatlist{ 
 \CCScat{K.6.1}{Management of Computing and Information Systems}%
{Project and People Management}{Life Cycle};
 \CCScat{K.7.m}{The Computing Profession}{Miscellaneous}{Ethics}
}

\teaser{
  \centering
  \includegraphics[width=\linewidth]{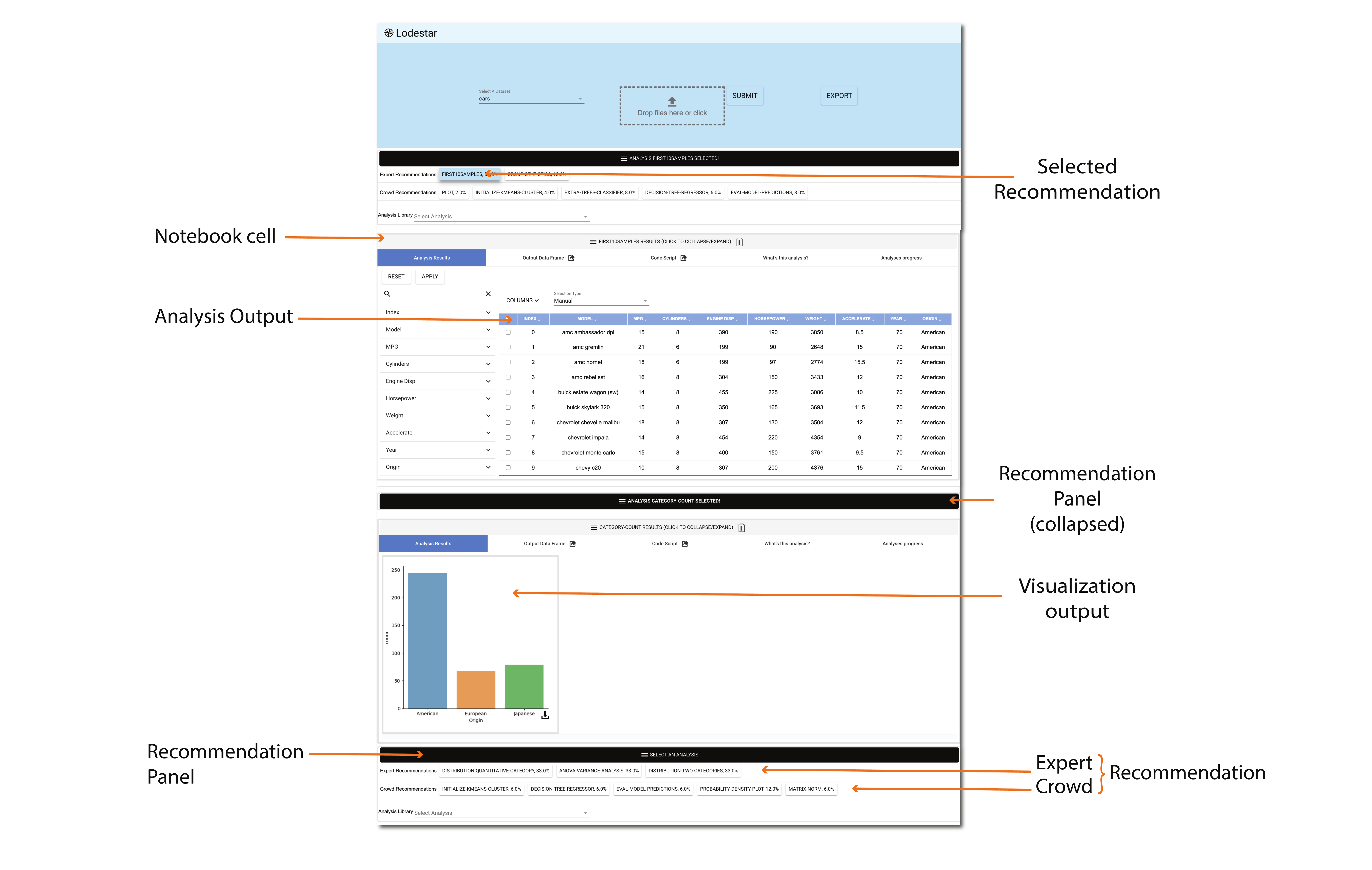}
  \caption{\textbf{Lodestar web interface.}
  The top panel, above the selected recommendations, provides a data selection menu.
  The black dividers between sections are recommendation panels combining suggested analysis steps from various sources (called \advisor{}s).
  Areas that have graphs and analysis outputted are analysis cells, each with multiple tabs:
  ``Analysis Results'' gives charts or tables, ``Output Dataframe'' and ``Code Script'' shows the outputs and current code \codeunit{}, and ``What's this analysis?'' gives a brief description of the analyses.
  Outputs, code, and charts can also be exported.}
  \label{fig:teaser}
}

\begin{document}

\maketitle

\section{Introduction}
\label{sec:introduction}

Data science is still a nascent and emerging discipline, which makes it challenging for analysts to learn and keep up with new tools and techniques.
There is already a dizzying array of libraries, such as Scikit-Learn~\cite{scikitlearn2011}, Pandas~\cite{pandas2011}, and TensorFlow~\cite{tensorflow2016}, and best practices and workflows change often.
Furthermore, few standardized methods exist for data analysis: many times, the exact data transformations, computations, and analyses needed depends on the data, task, and user.
This means that cookbook methods or simple workflow templates are insufficient to teach fledgling analysts how to tackle realistic and ever-changing data science problems.

In response, we present \textsc{Lodestar}, an interactive and visual sandbox environment for independent learning of new data analysis methods and best practices in data science.
Our aim in developing \sys{} is to simplify the process of finding and experimenting with new analysis methods by providing automated, data-driven recommendations.
We want \sys{} to be a self-contained environment for rapid learning and iteration, where everything the user needs to infer the function and purpose of an analysis step is available in one place.

The \sys{} system uses a computational notebook interface (similar to a Jupyter notebook~\cite{jupyternotebook}) showing a sequence of analysis steps in the form of Python code cells (see Figure~\ref{fig:teaser}), but enables the user to initially select from and interact with self-contained code cells without having to write any code.
The user merely selects which data frame to analyze, and the system displays a ranked list of recommendations of analysis steps to be executed on that data.
Each analysis step is represented by an interactive visualization in the notebook interface, giving the user insights into its output and behavior.
Furthermore, users can view the corresponding code for any analysis step, and even export the resulting notebook from \sys{}, providing flexibility in how users learn from and interact with \sys{}'s analysis recommendations.

\sys{} provides recommendations for the user's next analysis step based on the current state of the analysis and the dataset being analyzed.
Recommended analysis steps and workflows are derived from two sources representing current best practices in data science: (1) existing data science tutorials from online academies and training materials (i.e., an \textit{expert recommendation}), and (2) common analysis patterns mined from a large corpus of publicly available Jupyter notebooks~\cite{Rule2018} (i.e., a \textit{crowd recommendation}).
The code cells extracted from each source are manually curated, then programmatically clustered into synonymous analysis steps, and inserted into a large directed graph of connected cells representing common analysis workflows.
The \sys{} recommendation engine can then identify and rank the most relevant analysis steps given a specific position in the graph.


We developed \sys{} using an iterative design process. We used early feedback from six participants to improve the interface design. Our findings show that key \sys{} interactive features, such as automated recommendations, a visualization of the full analysis workflow, a code review pane for suggested analysis steps, and export support for Jupyter Notebooks, provide significant value to those who are learning data science.


In this work, we make the following contributions: (1) a holistic recommendation process involving two sources of data analysis practice: crowd-based and expert-based; (2) a sandbox interface design integrating visualizations, interactions, and code to facilitate learning about new data analysis techniques; (3) results from a formative study evaluating the \sys{} design; and (4) a unique data analysis architecture that integrates a recommender system with a computational notebook interface. 
All our materials, including source code, documentation, and study results, have been made available on the following OSF page: \url{https://osf.io/pztva/}

\section{Background}
\label{sec:background}

\subsection{Sensemaking}

\sys{} was architected to aid and encourage best practices in sensemaking. 
Richard Hamming described sensemaking as ``the process of searching for a representation and encoding data in that representation to answer task-specific questions''~\cite{Russell1993}.
Dubbed the \textit{sensemaking loop}~\cite{pirolli2005sensemaking}, each sensemaking iteration works to refine and build on the previous insights---ultimately enabling the analyst to address less specialized audiences~\cite{Thomas2005}. 
In combination, these iterations make up the data science workflow.
Analysts usually use visualizations or other types of intermediate results to motivate further analysis.
However, these results can sometimes be dead ends.
Kandel et al.\cite{kandel2011wrangler} found that analysts will overcome dead ends by backtracking and exploring new branches. 

\subsection{Interactive Visualization Design Environments}

Many visualization systems and toolkits are designed around specific data analysis tasks, making the analysis process easier to perform.
Excel supports basic visualization and data transformations.
Shelf-based visualization environments such as Tableau (n{\'e}e Polaris~\cite{stolte2002polaris}) allow easy configuration of visualizations through drag-and-drop of data attributes and metadata onto ``shelves'' representing visual channels.
This approach is flexible enough for even novice users to construct a wide range of visualizations.
Interactive visual design environments such as Lyra~\cite{Satyanarayan2014}, iVoLVER~\cite{Mendez2016}, and iVisDesigner~\cite{Ren2014} utilize direct manipulation to allow users to bind data to visual representations. 
More recently, Data-Driven Guides~\cite{Kim2017}, Data Illustrator~\cite{liu2018dataillustrator}, DataInk~\cite{Xia2018}, and Charticulator~\cite{Ren2019} provide advanced tools for representing data items as visual elements and mapping their attributes to data dimensions.
Keshif~\cite{Yalcin2017}, a faceted visualization tool, generates grids of predefined charts to support visual exploration by novice users.

Visualization development toolkits such as D3~\cite{bostock2011d3} and Protovis~\cite{Bostock2009} provide fine-grained control over designing interactive visualizations, but require significant programming expertise to use.
Visualization grammars, such as ggplot2~\cite{wickham2016ggplot2}, Vega~\cite{Satyanarayan2016_vega}, and Vega-Lite~\cite{vegalite}, abstract away implementation details, but still require programming knowledge to use.
Furthermore, even advanced visualization tools, toolkits, and grammars offer only limited functionality for manipulating the data, and only support a small number of statistical functions.

\subsection{Visualization Recommendation}

The purpose of visualization recommendation is to suggest relevant visualizations to the user to facilitate data analysis~\cite{herlocker2004collaborafilter}, where the visualizations are fully designed in advance and therefore directly accessible to the user.
It was first proposed by Mackinlay~\cite{mackinlay1986automating} in 1986 with automatic design of effective presentations based on input data.
The work combines expressiveness and effectiveness criteria from studies such as those by Bertin~\cite{bertin1983semiology} and Cleveland et al.~\cite{cleveland1984graphical} to recommend appropriate visualizations.
In 2007, Tableau's Show Me feature~\cite{mackinlay2007show} revealed a commercial product with the implementation of these ideas.
Following the idea of Mackinlay's automatic visualization, Roth et al.~\cite{roth1994interactive} enhances user-oriented design by completing and retrieving partial design graphics based on their appearance and data contents.
The rank-by-feature framework~\cite{seo2005rank} ranks histograms, scatterplots, and boxplots over 1D or 2D projections to find important features in multidimensional data.
SeeDB~\cite{seeDB2014} generates all possible visualizations given a query of the database and identifies interesting ones.
Perry et al.~\cite{perry2013vizdeck} as well as van den Elzen and van Wijk~\cite{van2013small} tackle the problem of generating small multiple visualizations shown as thumbnails using their statistical properties.

In the last few years, recommender systems have become widely used for visualization. 
Voyager~\cite{Wongsuphasawat2016} generates a large number of visualizations given a user-specified partial specification, and organizes them by data attributes.
The generated visualizations are rendered as cards on a scrolling view.
Saket et al.~\cite{Saket2017} propose the Visualization-by-Demonstration framework, which allows users to provide incremental changes to the visual representation.
The system recommends potential transformations such as data mapping, axes, and view specification transformations.
Zenvisage~\cite{siddiqui2016effortless} automatically identifies and recommends desired visualizations from a large dataset.
Voyager 2~\cite{Wongsuphasawat2017} extended the original Voyager through wildcard functionality that explores all possible combinations of attributes.
Most recently, Draco~\cite{Moritz2019} even automates visualization design itself using partial specifications and a database of design knowledge expressed as constraints. VizML learns what visualizations to recommend by training neural network models on millions of visualization designs made using Plotly~\cite{hu2019vizml}.

Several tools extend these ideas to recommending analytical insights and data processing steps.
``Top-K insights''~\cite{topkinsights} provides a theory for generating top $K$ insights from multidimensional data.
Similarly, Foresight~\cite{Demiralp2017} presents the top $K$ insights in a dataset from 12 insight classes using a corresponding visualization.
DataSite~\cite{cui2018datasite} organizes significant automatic findings in a specific feed of notifications.
Finally, Voder~\cite{Srinivasan2019} builds on a similar feed as DataSite to provide ``interactive data facts'' using visualizations.

Our proposed \sys{} system combines these ideas from visualization recommendation with an analytical perspective, and allows stringing together such analytical steps into a sequence. 
There are some existing efforts on recommending data analysis techniques and workflows.
Yan et al.~\cite{yan2020auto} demonstrate that online repositories of computational notebooks can be a valuable resource for modeling and testing a recommendation system for data cleaning techniques.
Milo et al.~\cite{bar2020automatically} take this a step further by automatically generating entire data exploratory workflows using deep reinforcement learning techniques.
Our system builds on these works by presenting a holistic model and code mining pipeline for deriving new recommendation features in a data-driven way, whether for data visualization, data preparation, or data analysis workflows.  Essentially, \sys{} extends the idea of automated recommendations to the entire data science pipeline, rather than visualizations only.

\subsection{Interactive Notebooks}

Donald Knuth's notion of a ``literate'' form of programming~\cite{knuth1984literate}, which merges source code with natural language and multimedia, has extended to the concept of \textit{literate computing} in the form of computational notebooks~\cite{kluyver2016jupyter, jupyternotebook, tabard2008individual}, that combine executable code, its output, and media objects in a single document.
This has proven to be very useful for rapid prototyping and exploration as well as for replicability and communication, particularly for data science and analysis~\cite{Rule2018}. 

Because of their success, with adoption even at the level of entire organizations~\cite{netflix}, notebooks have enjoyed significant progress in recent years.
The new generation of computational notebooks, such as Google Colaboratory~\cite{colaboratory} and Codestrates~\cite{radle2017codestrates}, enable synchronous collaboration.
Beyond such features, the JavaScript-based Observable notebook~\cite{observable} also supports one-way reactive execution flows. 

Visualization in particular has recently begun to adopt computational notebooks.
Altair~\cite{Altair} builds on Vega~\cite{Satyanarayan2014} and Vega-Lite~\cite{vegalite} to provide statistical visualizations in Python, and thus in Jupyter Notebooks as well.
Idyll~\cite{conlen2018idyll} supports a notebook-like markup language to create interactive data-driven document for communication.
Vistrates~\cite{Badam2019} provides a collaborative visualization workflow in a notebook.
Observable~\cite{observable} leverages the computational notebook environment to also provide a collaborative visualization platform. 
Literate visualization~\cite{Wood2019} integrates the visualization design process with the choices that led to the implementation. 

End-user and live programming paradigms have proven useful in creating intuitive interactions with visualizations found in computational notebooks.
For example, Wrex~\cite{wrexDrosos} and Mage~\cite{keryMage} leverage user interactions on data visualizations to automatically generate exemplar code.
As in traditional end-user programming platforms, Mage and Wrex demonstrate the link between code and visual interactions. 
Torii~\cite{headTorii} uses a live programming model to enable easy maintenance and reuse of source code, for the purposes of building tutorials.
These systems not only add to the number of ways users can interact with their literate document, but, they also create loops between code and visualization that lend well to data iteration.

\section{Motivating Scenario}

Here we describe a motivating data analysis scenario that highlights a knowledge gap among existing work that the \sys{} system aims to address.
Suppose that an undergraduate student has just started learning about data analysis techniques in a university data science course.
She is interested in selecting a dataset for her first assignment in the course.
However, before testing out any new analysis methods on an unfamiliar dataset, the user wants to start with the simpler case of analyzing the popular Cars dataset, which is made available via a drop-down menu in the \sys{} interface upon startup.
Upon selecting the Cars dataset, the user is presented with a list of initial data analysis steps, which have been recommended based on the user's choice of dataset (top of \autoref{fig:teaser}).
Each recommendation is represented as a clickable button.
As the user reads the labels of the recommendation buttons, she may position her mouse over each, triggering tooltips with a brief description of each recommendation's behavior.
The tooltip associated with the ``first 10 samples'' recommendation suggests that the user look at ``the first 10 rows of the data frame.''

\begin{figure*}[htb]
    \centering
    \resizebox{0.8\linewidth}{!}{\includegraphics{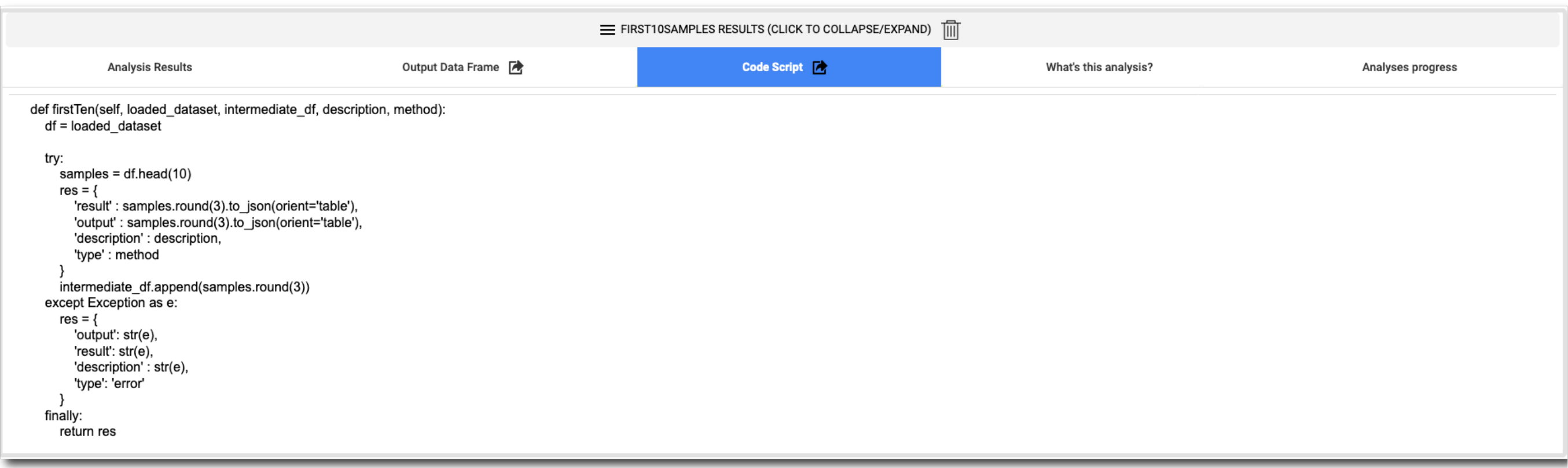}}
    \caption{Our fledgling analyst can explore the tabs within the analysis cell for details about the analysis step she chose to execute from the recommendation panel.}
    \label{fig:code-script}
\end{figure*}

To gain a better sense of what the Cars dataset contains, the user decides to select the ``first 10 samples'' recommendation.
In response, \sys{} adds a notebook cell to the interface.
This notebook cell contains results comprised of the Python code used to compute the analysis step (Figure~\ref{fig:code-script}), a brief description of the analysis technique, and a visualization of the results, in this case a table displaying the first 10 rows of the cars dataset (middle of Figure~\ref{fig:teaser}).
The table shows that the Cars dataset contains 10 attributes, where at least three attributes appear to be categorical. The user is curious about what these categorical attributes contain.

\sys{} generates a new panel of recommendations below the new analysis cell, allowing the user to select the next step of her analysis workflow.
In this round of recommendations, she sees a recommendation for generating ``group statistics,'' which, according to the associated tooltip, promises to generate descriptive statistics of the dataset.
These statistics could help her better identify the categorical attributes.
Upon selecting ``group statistics,'' another notebook cell is added to the interface, along with a third recommendation panel with potential follow-up analyses (bottom of Figure~\ref{fig:teaser}, collapsed).

In the table of summary statistics produced by the ``group statistics'' recommendation, the user notices that there are three unique categories in the ``Origin'' attribute.
Curious about the distribution of cars per country of origin, she scans the previous round of recommendations for one that will allow her to dive deeper into categorical attributes.
She finds the ``category count'' option within the second recommendation panel from which she originally chose ``group statistics,'' where ``Category count'' will show ``the distribution of categorical attributes across different values.''

When she updates her selection to be ``category count'' instead of ``group statistics,'' the original notebook cell for ``group statistics'' is removed and replaced with a new notebook cell displaying the results of ``category count''; the third panel of recommendations is also updated accordingly.
The output of the ``category count'' cell is a bar chart showing the number of cars for each country from the Origin attribute (bottom of Figure~\ref{fig:teaser}).
The user observes that there are more ``American'' cars than any other kind.

The user now knows more about both the dataset and how to implement some of the statistical techniques she had encountered in class.
She is curious to explore the Python code she observed in the notebook cells and the characteristics of cars from different countries.
She scrolls up to the menu panel at the top of the interface and selects ``export notebook''to save her current workflow as a Jupyter Notebook file  (\texttt{cars\_analysis.ipynb}) for further analysis and manual editing.

\section{Design Requirements}
\label{sec:design-requirements}

Our goal is to make \sys{} an interactive and visual sandbox environment for learning and experimenting with new data science methods in a data-driven way.
We also wanted to make data science universally accessible to fledgling data analysts and enthusiasts alike.
These core ideas helped us compile a set of design requirements and some preliminary prototypes.
In this section, we outline our major design requirements, and report on a formative study conducted to validate and refine our approach to the \sys{} interface design and system development processes.

\begin{itemize}[nosep]
\item\textbf{D1: Informed by best practices.} Generated recommendations should be drawn from current practice, empowering those new to data science to learn how to effectively analyze their data~\cite{guoTeach, mackinlay2007show}.

\item\textbf{D2: Prioritize analysis steps over code.} Our intended users are trying to analyze data in a fast and fluid fashion, but may not yet be familiar with specific libraries or modules needed to complete different analysis steps.
\sys{} needs to build a bridge between the high level analysis steps common in data science, and the low-level code needed to accomplish these steps~\cite{insideInsights}.
For example, recommendations should be immediately relevant and situated within the overall data science pipeline to enable users to progress in their analysis.

\item\textbf{D3: Enable independent exploration.} To ensure that users can explore their data independently, educational interface elements must also be incorporated to automatically provide documentation and clarification of system behavior~\cite{trustAutoAI}. Furthermore, intermediate and final results should be presented using visual representations that can be easily interpreted regardless of user expertise~\cite{tufte2001visual}.

\end{itemize}

\subsection{Formative Study}
\label{sec:design-framework}

We conducted a formative user study to evaluate the usability of an early prototype of the \sys{} system, which we used to validate our initial design requirements and refine the system design. This study was approved by our home institution's IRB.

\subsubsection{Study Design}

The study was conducted over a period of one month in which we interviewed 6 fledgling analysts and data scientists; all undergraduate university students.
We focused on recruiting university students, since they are generally learning data science methods for the first time and thus could provide helpful insights in our design process.
Each student had demonstrated knowledge of data science fundamentals through attending a university-level introductory data science course and/or other relevant machine learning/data science experience.
Although not a prerequisite of the recruitment process, some students also had experience performing analysis on platforms such as Excel and Tableau.

\subsubsection{Method}

Each interview lasted for 60 minutes and was divided into three phases.
Prior to the interview, each participant signed a consent form, allowing us to record audio and screen capture throughout the duration of the interview.
The first phase consisted of questions, delivered verbally, that assessed the participant’s recent experience in learning data science techniques and tools through classes, side projects, research, and other such activities.
During this section, participants were also asked specific questions regarding their view on recommender systems.

\textit{The second phase} of the interview was dedicated to introducing an early prototype of the \sys{} system in which participants were given a brief 2-minute description of Lodestar and associated goals.
The next 5 minutes were spent giving the participant a cursory tutorial of the system.
For each participant, the tutorial was given using a pre-written script and with the same sample data set to give each of them equal knowledge of the system prior to their exploration.
The participants then spent the next 15-20 minutes using the \sys{} system to conduct exploratory data analysis on a data set of their choosing.
We restricted their choices to two datasets; the Boston House dataset from a Udacity tutorial,\footnote{The Udacity tutorial is available here: \url{https://github.com/sajal2692/data-science-portfolio/blob/master/boston_housing/boston_housing.ipynb}.} and the ubiquitous Cars dataset.
During this exploratory session, participants verbalized their thought process, questions, and comments with a think-aloud protocol.
We encouraged participants to ``to use any and all features of the \sys{} system'' and to ``explore whatever aspects of the data [they found] interesting.''
Participants were allowed to end the session before the allotted time expired if they were satisfied with their results.

\textit{The third and final phase} of the interview consisted of a post-exploration questionnaire that asked participants to describe the utility of \sys{} for their common data analysis tasks.
They were specifically asked if they would adopt \sys{} to learn new data science techniques and whether they trusted the recommendations. 

All sessions were held in a lab environment using Google Chrome on a Macbook Pro with a 15-inch Retina display.
Audio was recorded using the built-in voice recording application on a mobile device.
Screen capture was done using Apple's QuickTime Player.
Observational notes from the study coordinators, text responses from our questionnaires, and audio and video recordings were collected for further analysis and prioritization of design requirements and functional features of the existing prototype. 

\subsubsection{Results}

Our formative study found that a majority of participants were in favor of using \sys{} in their daily work, but suggested several modifications to make the system more useful.
For the sake of brevity, we focus primarily on summarizing their constructive feedback below (participant IDs start with ``FP''): 

\paragraph{Provide Clear Documentation \& Context}

Our early prototypes did not include tooltips or descriptions of analysis steps.
Several participants highlighted the need for increased transparency in the interface. 
Specifically, they wanted clearer naming conventions, documentation of features and methodologies (e.g., the difference between expert and crowd recommendations), and explanation of expected system behavior.
For example, some participants had difficulties understanding the meaning of certain user interface elements.
Participants asked questions such as ``\textit{what are these percentages?}'' (FP6), or ``\textit{[what do] the columns on the left side represent?}'' (FP5).
Participants FP1, FP2, FP3, and FP5 also asked if there ``\textit{is actually a way to view the entire dataset?}'' (FP2).

There were many questions specific to the meaning of recommendations.
For example, FP3 said ``\textit{I think the names [are] misleading... there were some really complicated names for just a simple linear regression. [It] should just be changed [to more] obvious names.}''
Similarly, FP2 suggested that there should be ``\textit{a longer description [...] [or] some way to show their effectiveness without the user having to Google search them.}''
These misconceptions indicate that better documentation is needed to help new users understand the interface.

\paragraph{Improve Tracking of Analysis Progress}

Several participants wanted to be able to see what phase of the data science process they were in based on the current state of their analysis workflow. Our early prototypes did not include the feature to track previously selected analysis. 
FP4 drew parallels with a restaurant order tracker, where \sys{} should partition each part of the data science process into separate steps, and group analysis recommendations into these steps.
Users would then be able to better understand their progress within the data science process.

\paragraph{Enable More Granular Control}

The early \sys{} prototype only allowed users to choose from pre-loaded datasets, and did not provide any export or customization functionality for analysis steps.
However, multiple participants expressed the desire to import their own dataset and export their own code for later sharing and reuse.
Participant FP4 said that they would be frustrated if they wanted to ``\textit{export it or make some changes in the data or [try] to do something that is not supported by Lodestar [while] not having any way of doing so.}''
Participants also highlighted the need for more control over what parameters or attributes were being passed into different analysis steps, such as selecting specific attributes when generating visualizations or executing regression analyses.
These observations suggest that users should be able to import their own dataset, customize analysis steps, and export their current analysis workflows.

\subsubsection{Further Refinement of \sys{}}

Though participants could see promise in providing automated recommendations (design requirement \textbf{D1}), the expressed need for more tracking of workflow structure and progress also reinforces design requirement \textbf{D2}.
Without additional context to help users situate themselves within the broader data science process, users can easily lose their train of thought, hindering their analytic flow.
The need for more documentation and control observed in our formative study supports design requirement \textbf{D3}.
Without adequate information, users are unable to explore new data analysis techniques and interpret the results in \sys{} on their own.
Users also find it difficult to tailor their explorations to their specific needs without access to the code.

These points of feedback served as motivation for additional iteration on the \sys{} feature design.
Specific features that were added as a result of this study included the ability to export the user's notebook to an \texttt{.ipynb} file for use outside of the system, a visual tracker that displays the progress of the user's analysis in each output cell, showing which recommendations have been chosen so far, and descriptive tool-tips of the different analysis techniques in each output cell.

\begin{figure*}[htb]
    \centering
    \includegraphics[width=0.8\linewidth]{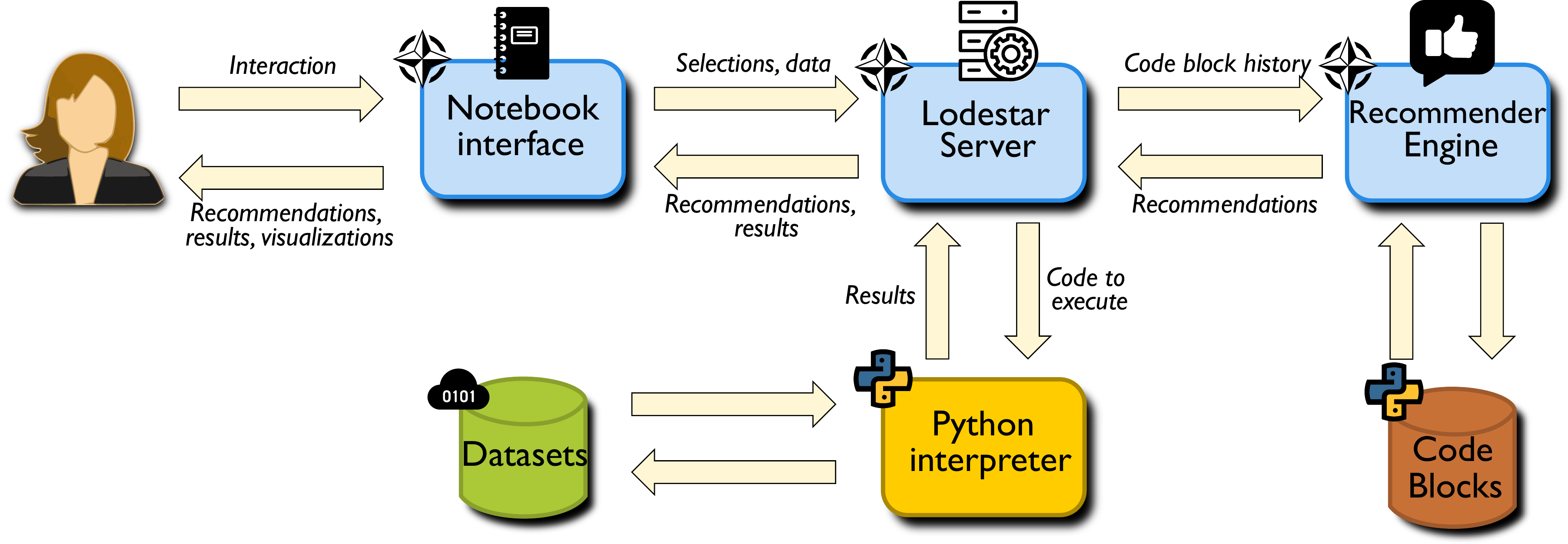}
    \caption{Overview of the Lodestar architecture. The user interacts with the notebook interface and selects either a data set to bootstrap the notebook or an analysis step within a guided workflow. The notebook interface sends the selection as a request to the Lodestar server. The Lodestar server sends requests to the recommendation engine for subsequent recommendations based on current selections (data or analysis). Lodestar server also sends a request to the Python interpreter to execute any selected analysis. Results from both these requests are sent back from the Lodestar server to the notebook interface for the user to view and interact.}
    \label{fig:lodestar-overview}
\end{figure*}
\section{System Overview}
\label{sec:overview}

\textsc{\sys{}} is \textit{data analysis recommender}, i.e., a system that interactively suggests the next step to take in an analysis workflow.
\sys{} is designed in the style of an interactive computational notebook, and generally inspired by the designs of existing notebooks such as Jupyter~\cite{jupyternotebook}, Observable~\cite{observable}, and Google Colaboratory~\cite{colaboratory}.
Given Python's broad popularity in data science contexts~\cite{Rule2018}, we chose to focus on Python as our target computation environment.

\sys{} consists of four main components, shown in \autoref{fig:lodestar-overview}: a browser-based \textit{notebook interface}, an \textit{interactive computing protocol}, a \textit{recommendation engine} to suggest analysis steps, and a server-side \textit{kernel}~\cite{jupyternotebook} to execute analysis steps.
The protocol manages communication between the client and server (commands as well as computational results), and the kernel on the server side runs each analysis step that the user selects using an interpreter.
The Flask server handles all of the client requests for data processing, analysis, and recommendations, with different endpoints.

\sys{} emphasizes an iterative workflow design where analysis steps are added progressively, one at a time, providing finer-grained control to the user.
To help users focus more on analysis steps and best practices rather than low-level code, \sys{} allows the user to rapidly choose from a list of recommended analysis steps. 
These recommendations are displayed in the form of buttons, so a user can easily select and execute an analysis step of interest with a single click.
Furthermore, these recommendations are mined from recent Python tutorials and active GitHub repositories of Jupyter Notebooks, enabling the user to construct new analysis workflows based on best practices in a data-driven way.

We describe the \sys{} notebook interface in more detail in \autoref{sec:notebook-interface}.
We describe our process for extracting, curating, and recommending analysis steps in \autoref{sec:advisors-and-recommendations}.

\section{Notebook Interface}
\label{sec:notebook-interface}

The \sys{} interface (shown in \autoref{fig:teaser}) is an interactive notebook providing a literate computing environment~\cite{knuth1984literate} that runs in a web browser on the client. 
Similar to existing notebooks, the \sys{} notebook is essentially a linear document that the user can selectively edit and execute.
The interface contains three major components: a menu panel at the top, one or more notebook cells, and recommendation panels for each cell.
The notebook cells and recommendation panels dynamically appear and update within the notebook interface in response to user interactions. 

The user begins their analysis using the menu panel to load an existing dataset or a new dataset (in CSV format) into the system.
Once a dataset has been loaded, \sys{} generates a recommendation panel within the notebook interface, providing the user with an initial set of recommended analysis steps.
We refer to the actual code behind each analysis step as an \emph{analysis \codeunit{}}, and the displayed result of executing the analysis step as a \emph{notebook cell}.
From this point onward, the analysis process forms a cycle that repeats until the user is satisfied with their new workflow:

\begin{enumerate}[nosep]
    \item The user \textbf{selects} an analysis step from a \textit{recommendation panel}; 
    \item The kernel \textbf{executes the matching analysis block} on the server;
    \item The notebook \textbf{displays the output} by appending a new cell; and
    \item The notebook \textbf{generates a new panel of recommendations}, based on the user's previous selection.
\end{enumerate}

When the user is ready to migrate their workflow to a complementary tool, for example to iterate on the code directly within a code editor, they can export the \sys{} workflow as a Jupyter notebook file.

\subsection{Recommendation Panel}
\label{sec:recpanel}

Every notebook cell in the \sys{} interface has an accompanying recommendation panel, allowing the user to extend their latest analysis step by one cell.
When the user selects an analysis step from a recommendation panel, a new notebook cell is generated for the selected recommendation, along with a new recommendation panel underneath.
\sys{} uses the output of the preceding notebook cell as the input for executing any analysis step selected in this recommendation panel.
Each panel provides two sets of recommendations, one from a \textit{crowd advisor} and one from an \textit{expert advisor}.
The crowd advisor sources recommendations from online data analysis repositories such as GitHub.
The expert advisor sources recommendations from educational resources such as textbooks, online classes or online tutorials. We describe the \sys{} advisors in \autoref{sec:advisors-and-recommendations}.

If a user is unsatisfied with a given set of recommendations, they can choose from \sys{}'s full catalog of analysis steps in a drop-down menu at the bottom of each recommendation panel. 
This list is available in the supplementary materials.

\subsection{Notebook Cell}
\label{sec:nbcell}

Once a selection is made in a recommendation panel, the selected analysis step is highlighted and the results are displayed in a new notebook cell, allowing the user to review their past selections and the corresponding results with each subsequent step.
Furthermore, the user is able to go back and update the results at any time by selecting a different analysis step in any of the previous recommendation panels.
Any cell can also be deleted, which triggers the removal of all downstream cells that depend on the deleted cell.
In this way, \sys{} maintains a linear structure in the notebook, making it easier for users to navigate within the analysis workflow.

To help users understand the functionality of each recommended analysis step and its purpose within the context of the larger data science process, notebook cells consist of five tabs.
Each tab describes the behavior of the analysis block represented by this notebook cell.
We refined the design of each tab based on the feedback we received from the formative study (see ~\autoref{sec:design-framework}):

\begin{itemize}[nosep]
    \item\textbf{Output Data Frame:} Default view that renders the output data frame produced by executing the analysis step as a table.
    \item\textbf{Analysis Results:} Displays the raw results produced by the analysis step  (e.g., a \texttt{print} statement, or Seaborn visualization).
    \item\textbf{Script:} Displays the Python code within the corresponding analysis block.
    \item\textbf{``What's this Analysis?'':} Provides a brief, high-level description of the analysis step.
    \item\textbf{Analysis Progress:} Displays the chain of analyses leading to the current analysis step, where each step has an intuitive name.
\end{itemize}

\subsection{Exporting Code and Results}
\label{sec:export}

When the user is ready to migrate their analysis workflow to a related tool, they can export content directly from \sys{}.
To export the code for a specific analysis step into an independent Jupyter notebook file, the user can click on the export button next to the \textit{Code Script} tab of the corresponding cell.
To export the entire analysis workflow, the user can click on the export button on the menu panel at the top of the interface.
Similarly, \sys{} enables users to export the output data of any displayed notebook cell in the form of a CSV file.
To do this, the user clicks on the export button next to the \textit{Output Data Frame} tab.
The user can also download the visualizations displayed in any notebook cell as separate PNG files.

\section{\Advisor{}s and Recommendations}
\label{sec:advisors-and-recommendations}

The \sys{} recommendation engine is based on the notion of an \textit{\advisor{}}: a source of analysis recommendations.
\sys{} supports multiple \advisor{}s, each consisting of a \codeunitcontainer{} of analysis steps and a set of advisor-recommended transitions between analysis steps (i.e., a recommendation graph).
In our current implementation, we use two \advisor{}s: a ``crowd'' \advisor{} drawn from our semi-automatic code analysis, 
and an ``expert'' \advisor{} drawn from the manual code curation. 
For each advisor, the recommendation panel will show a list of up to five recommendations, ordered by probability, or how frequently this analysis step came next in the respective recommendation graph.

In this section, we describe how we build our recommendation graphs for the expert and crowd \advisor{}s, and how we enable \sys{} to identify equivalent or related states across both graphs.

\begin{figure}[htb]
    \centering
    \includegraphics[width=\columnwidth]{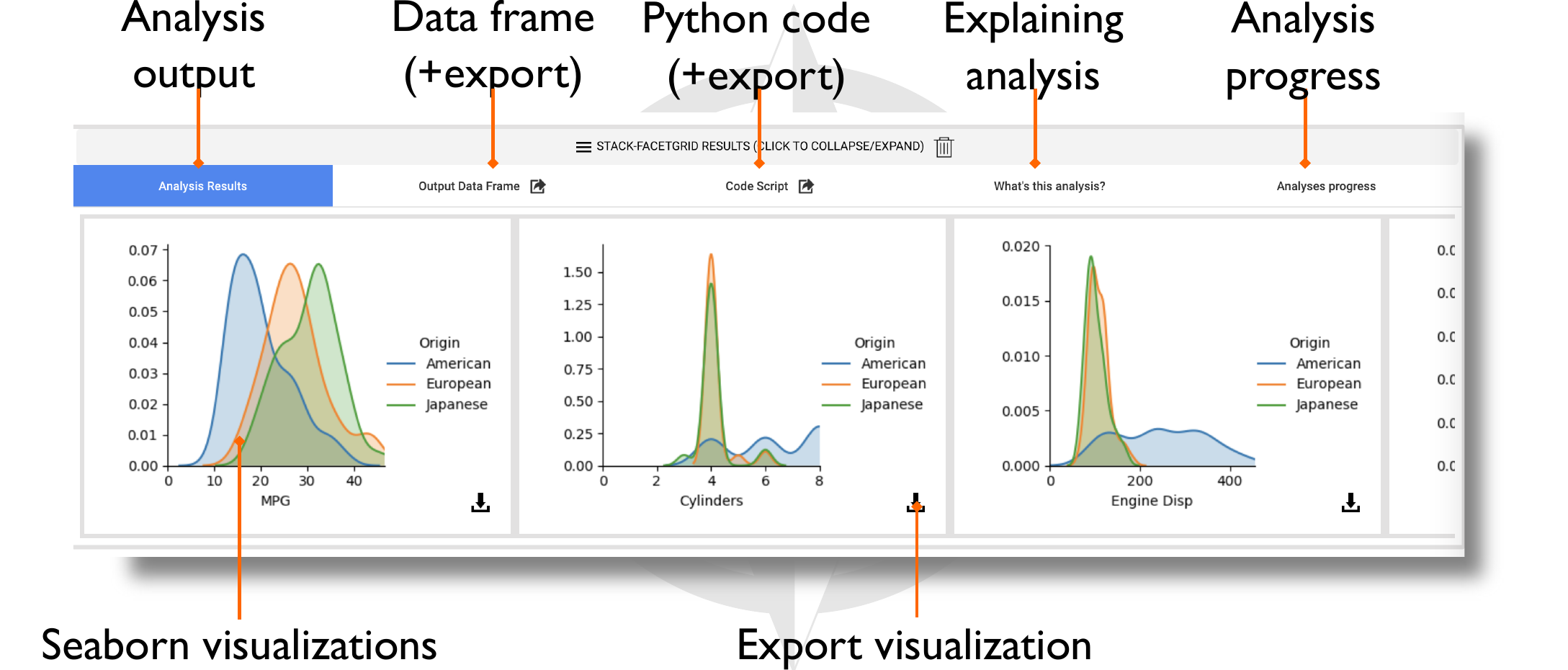}
    \caption{Figure grid generated by an analysis block using the Seaborn statistical data visualization package for Python.}
    \label{fig:figure-grid}
\end{figure}

\subsection{Recommendation Graph}

\sys{} models transitions between analysis steps by treating analysis workflows (e.g., existing tutorials or computational notebooks) as paths taken through a network graph. Each node in the graph is an analysis step, and a directed edge appears in the graph for each pair of consecutive analysis steps observed in a workflow. \sys{} leverages the relative frequency of these transitions to predict which analysis steps are likely to occur next.
The particular graph structure used in \sys{} is a Markov chain, and the final computed graph we refer to as a \textit{recommendation graph}.

\sys{} traverses the recommendation graph one state at a time for each user input (i.e., choice of analysis step).
As a result, our recommendation approach does not require maintaining specific state about the analysis itself.
Instead, the location in the Markov chain serves as state, and transitions (e.g., recommendations) thus depend only on the current state.

\begin{figure}[htb]
  \centering
  \includegraphics[width=\columnwidth]{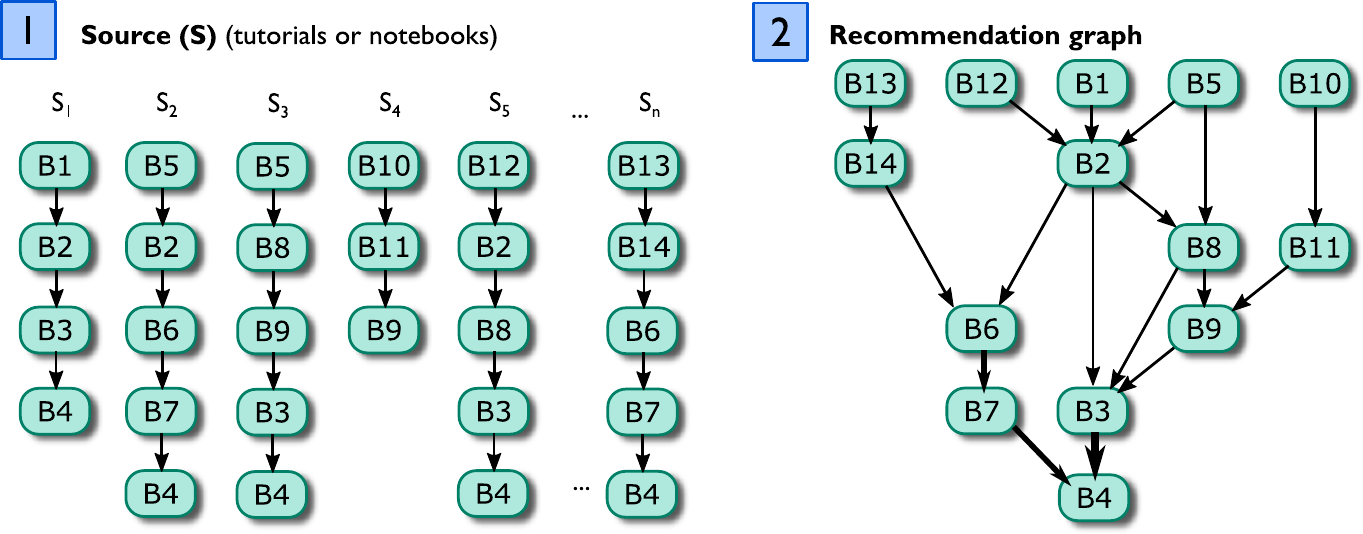}
  \caption{Mining \codeunit{}s into a recommendation graph representing a Markov chain.
  In Step 1 (left), sources $S_1, \ldots, S_n$ (manually curated or automatically extracted) yield (ordered) sequences of \codeunit{}s $S_i = (B_1, \ldots, B_m)$.
  In Step 2, a recommendation graph can be derived by matching \codeunit{}s that appear in multiple sequences and joining the sequences at those nodes.
  Edges between \codeunit{}s in the graph are the frequency-weighted state transitions in the chain.
  }
  \label{fig:reco-graph}
\end{figure}

We can infer these recommendation graphs programmatically by mining analysis \codeunit{}s (i.e., code snippets) from existing computational notebooks.
In this case, the analysis \codeunit{}s are used as the graph states, in place of their corresponding analysis steps.
~\autoref{fig:reco-graph} shows the general approach for mining analysis \codeunit{}s into this recommendation graph.
We extract the analysis \codeunit{}s from existing computational notebooks and recover the transitions between states from the sequences observed in each notebook, with the weights signifying the frequency of observed transitions.
Analysis \codeunit{}s become nodes $B_i$ in this graph, and edges represent probabilistic transitions $Pr(j|i) = P_{i,j}$, where the probabilities $P_{i, j}$ are taken from a stochastic matrix $\mathbb{P}$ that simply represents the frequency of transitions between \codeunit{}s in the individual sequences.

To infer the full recommendation graph, we first construct a separate Markov chain for each notebook (or tutorial) identified as a source for our \advisor{}s. 
Specifically, we model each notebook as a Markov chain with one state per \codeunit{} and the transition probability to move from \codeunit{} $B_i$ to the next \codeunit{} $B_{i+1}$ for each time step (e.g., user input) expressed as $Pr(i + 1|i) = 1$.
Similar analysis steps are labeled with the same high-level identifier, representing a broader category of computation that transcends individual notebooks (e.g., $B1$, $B2$, etc. in \autoref{fig:reco-graph}).
The result is a larger two-dimensional nested list, where each notebook is one row within the list (i.e., the left side of \autoref{fig:reco-graph}), and each column a sequence of analysis step categories.

We can then merge the resulting sequences into a single graph  (e.g., merging $S_1$, $S_2$, etc.\ in \autoref{fig:reco-graph}), and aggregate the relative frequencies associated with the different categories to determine transition weights (i.e., how often do we see \codeunit{}s from category B1 executed before \codeunit{}s from category B2?). 

Specifically, the transition probability $P_{i, j}$ (and thus edge weight in the recommendation graph) for the $i^{th}$ row and the $j^{th}$ column is the number of edges from $B_i$ to $B_j$ across all the sequences, divided by the out-degree of $B_i$.
In other words, the graph will have no edges (weight 0) between \codeunit{}s that never appeared in sequence, and will have normalized weights for \codeunit{}s that fan out to multiple different destinations (because they are used by many notebooks).
To bootstrap the recommendation, we recommend the first analysis in all the sequences (the root nodes in the graph).

\subsection{Extracting Analysis \Codeunit{}s for the Expert \Advisor{}}
\label{sec:extracting:expert}
We extracted analysis \codeunit{}s for our expert \advisor{} from online tutorials\footnote{Please see our supplemental materials for a detailed report on our full process for extracting and curating analysis \codeunit{}s for the expert \advisor{}: \url{https://osf.io/3gpsy/}}.
These tutorials were either Jupyter Notebooks or blogs which clearly delineated code from text.
Analysis \codeunit{}s correspond directly to code cells found in tutorial notebooks, or self-contained code snippets found in blog posts. 

While there exist many data science resources online, their focus and depth varies widely, from simple hands-on learning for beginners (e.g., software installation, basic Python knowledge, and Jupyter functionality), to expert-level guides on deep learning, sensitivity analysis, and model building and tuning. 
As a rule, we picked individual resources focused on teaching how to complete a specific analysis task.

We narrowed our search to end-to-end data science examples, which provide concrete sequences of analysis steps along the data science pipeline.
Specifically, we selected examples that have: an explicit purpose for the data analysis, step-by-step explanations and results, and runnable code.
These requirements helped to ensure that the extracted analysis \codeunit{}s will have similar functionality across examples.

\subsection{Formatting Analysis \Codeunit{}s for the Expert \Advisor{}}
\label{sec:formatting:expert}

To ensure that the extracted analysis \codeunit{}s are executable in \sys{}, we also apply a separate code curation process.
From our experience, each source has a specific analysis goal, and the \codeunit{}s across different sources may use different libraries, data attributes, and variables to achieve it.
For example, a tutorial using the Boston housing dataset, may generate a scatter to examine a linear relationship between four housing attributes, while in a school test-scores dataset it only makes sense to examine a linear relationship in between two attributes.
This is useful nuance for manual analysis, but cannot be directly used in a generic data analysis system such as \sys{}.
In other words, the analysis \codeunit{}s must be curated---typically generalized---to be applicable across multiple applications.

The \codeunit{} curation process is idiosyncratic, but consists of the following steps:
(1) adding missing dependencies, (2) replacing data-specific labels and attributes, (3) setting appropriate default parameters, and (4) generalizing code to operate on general data frames and output data frames too.
This process is very similar to our curation strategy for recommendations from our ``crowd'' \advisor{}.
We manually compared new \codeunit{}s to exiting \codeunit{}s within the \codeunitcontainer{}, to ensure there were no duplicates. 
Upon completion of the curation process, each new analysis \codeunit{} is added to the \codeunitcontainer{} for the recommendation graph.

\subsection{Managing Analysis \Codeunit{}s for the Crowd \Advisor{}}
\label{sec:extracting:crowd}

We extracted analysis \codeunit{}s for our crowd \advisor{} from a corpus of approximately 6,000 Jupyter notebooks, originally collected by Rule et al.~\cite{Rule2018}\footnote{Please see our supplemental materials for a detailed report on our full process for extracting and curating analysis \codeunit{}s for the crowd \advisor{}.}.
We filtered out notebooks which did not contain
import statements and API calls using common data science libraries, such as Numpy~\cite{van2011numpy}, Scikit-Learn~\cite{scikitlearn2011}, or Pandas~\cite{pandas2011}.
We first partition each notebook into discrete analysis \codeunit{}s. 
For Jupyter notebooks, the code is often already partitioned by the notebook authors through the use of Jupyter notebook code \emph{cells}.
Our straightforward approach is to identify existing cells in the Jupyter notebook corpus as separate analysis \codeunit{}s for \sys{}.

Our key insight for this process is that \emph{similar data analysis steps often use similar API calls} in the code.
Using this idea, we construct a term vector to represent each analysis \codeunit{}, where the vector represents the normalized frequency of each API call that appears within the \codeunit{}.
Each cell in the vector represents a unique API call observed in \emph{any} notebook in the dataset, allowing the vectors for any analysis \codeunit{} to be compared with any other \codeunit{} in the dataset.

We use these term vectors to cluster the analysis \codeunit{}s.
Specifically, the normalized vectors are passed to a $k$-means clustering algorithm to be clustered for similarity.
After some iteration, we identified 200 clusters as an ideal number for grouping the analysis \codeunit{}s extracted from our corpus (please see our supplemental materials for more details). 
Each resulting cluster represents a set of analysis \codeunit{}s that share similarities in functionality, and thus could also represent a shared or synonymous analysis step across the corresponding Jupyter notebooks.

Of the 200 representatives (one for each cluster), we ultimately selected $22$ \codeunit{}s as a starting set for the \sys{} \codeunitcontainer{}. 
For any given cluster, \sys{} needs a way of recommending a single analysis \codeunit{} to users that represents the corresponding analysis step.
We use code-line count as a heuristic to pick a representative analysis \codeunit from each cluster.
Specifically, we pick the \codeunit{}s which have a median number of lines relative to all other \codeunit{}s within a cluster. 

\Codeunit{}s for both the crowd and expert \advisor{}s are formatted to follow the same consistent structure assumed by the \sys{} system.
We format each analysis \codeunit{} to be a Python function, include necessary imports, convert the function's input and output to a data frame, and remove print statements and irrelevant comments.

\subsection{Identifying Synonymous States Across Advisors}

Of course, managing multiple \advisor{}s means that the system must track the state of the analysis in the recommendation graph for \emph{all} \advisor{}s when the user selects a recommendation from a specific advisor.
Our current solution uses a multi-level tagging mechanism where each \codeunit{} is manually tagged given its functionality; for example, a decision tree \codeunit{} could be tagged with \texttt{train-model} and \texttt{test-model}. 
Tags correspond to steps in the data analysis workflow.
We developed an understanding of these steps using previous studies~\cite{kandel2011wrangler, heer2008graphical, battle2019characterizing, zgraggen2018investigating, yan2020auto}.
Much like Yan et al.~\cite{yan2020auto}, we cast particular Python APIs to specific analysis steps.
For example, Pandas \textit{dropna} function was cast as a data-cleaning operation since dropping empty elements is a common way to clean data.
Our tags include:  \texttt{statistical-sampling}, \texttt{visualization}, \texttt{data-organization}, \texttt{data-cleaning}, \texttt{data-formatting} and \texttt{statistical-summary}.

In tagging analysis in this way, we allow for matching the new state of the specific \advisor{}, chosen by the user, to relevant states in the other \advisor{}s.
More specifically, if the user chooses a recommendation from the expert \advisor{} that suggests running a specific decision tree \codeunit{}, the \sys{} engine will advance the crowd \advisor{} to a state in its recommendation graph that corresponds to the \texttt{train-model} and \texttt{test-model} tags.
This design, as well as ordering recommendations by probability ordering, allows \sys() to guide best practices. 

The same functionality is used when the user eschews all of the recommendations and instead selects directly from the \codeunitcontainer{} through the drop-down box in the recommendation panel.
In this case, all of the \advisor{} models will be advanced to the appropriate state matching the \codeunit{} that the user executed.
This allows the user to iterate and sandbox different techniques, unhindered by a guided system. Though, this is the limit to manual user control that \sys{} supports.
\section{Discussion}

We have presented \sys{}, a computational notebook for rapid experimentation and learning of new data science practices.
Instead of forcing fledgling analysts to search for and apply relevant data analysis methods by hand,
\sys{} recommends suitable next steps for the current workflow using both manually curated as well as automatically crowd-sourced guidance.
Our work on \sys{} has uncovered several interesting discussion points: the prospect for data science for novices, the actual ``wisdom'' of crowd recommendations, and alternate recommendation mechanisms.

\subsection{Data Science for Non-Experts}

The real power of \sys{} lies not in its data sources, which are publicly available to anyone online, but in its ability to synthesize the knowledge from these diverse sources into a single unified model.
By sharing this knowledge in the form that data scientists are most familiar---Python (or R) source code---\sys{} provides reusable building blocks that can easily be transferred across data science workflows.

However, for the tool to be truly effective for its purpose, the \codeunitcontainer{} of analysis \codeunit{}s must be expanded and drawn from a large set of sources.
For example, new data sources could be incorporated to customize \sys{} for specific disciplines such as bio-informatics, computational journalism, and computer vision.
\sys{}'s advisor model may be one way to support this; instead of the ``expert'' vs.\ ``crowd'' dichotomy that our current implementation uses, a more robust implementation could support a plethora of pluggable advisors drawn from a central repository.
In this way, the advisors, analysis \codeunit{}s, and \codeunitcontainer{} could be community-driven and improved by anyone.

Choosing an analysis step or interpreting results in our current prototype still requires baseline data science knowledge, such as from a university data science course (indeed, all our participants had this).
However, the \sys{} approach does alleviate lack of \emph{expertise} in data science practice, which is often the case for academic learning.

\subsection{On the ``Wisdom of the Crowd'' for Data Analysis}

While we are excited about the prospects of the ``wisdom of the crowd''~\cite{Surowiecki2004} for data science and analysis, it has become clear that this is an area that will require significantly more work. 
For example, our current approach is not entirely automated; manual curation is still required in choosing a representative \codeunit{} from the clustering analysis and in editing the \codeunit{} into the appropriate form that \sys{} expects, including eliminating side effects, removing output statements, and resolving dependencies.
We plan to automate these steps in the future.

The need for manual curation, or at least review, is exacerbated by the fact that a significant portion of the code we analyzed in Rule et al.'s Jupyter notebook corpus~\cite{Rule2018} was of low quality: some notebooks had cells with a single line of code, or all of the source code in a single cell.
Many had non-functional code, syntax errors, or code that was never used.
While we have filtered these notebooks from our analysis, the signal-to-noise ratio in crowdsourced code is often low. 

The remedy for many of these challenges can often be found in sheer scale. 
While we studied the ``sampler'' dataset containing 6,530 notebooks in this paper, the full 600 GB dataset contains more than 1.25 million notebooks. 
With access to this many examples, we could afford to discard more problematic ones.
Furthermore, frequency of use would help ensure that best practices are easier to identify.
Of course, a dataset of this size brings with it a new set of scalability challenges.
Existing data processing~\cite{mudgal2018deep} and code analysis~\cite{glassman2015overcode, glassman2018visualizing} techniques could help address this big data challenge in the future.

\subsection{Different Recommendation Strategies}

The \sys{} recommendation engine is based on Markov chains, which are useful for representing a sequence of chained states or commands, as in a data science script.
However Markov chains may oversimplify the relationships between analysis steps and data science users in some ways.
It would be interesting to study how to use more sophisticated methods as part of the \sys{} recommendation engine.
For example, state-of-the-art recommender systems tend to be organized into collaborative filtering, content-based filtering, and hybrid filtering~\cite{Adomavicius2005}.
Collaborative filtering is based on a social view of recommendation, where behavior by other users such as navigation, ratings, and their personal traits are used to match content to a specific user.
In the case of \sys{}, this would enable the historical preferences of \sys{} users to guide other users.
For content-based filtering, recommendations can be derived by comparing items to recommend with user preferences and auxiliary information.
This approach could enable \sys{} users to be matched to specific analysis steps based on, e.g., workflows they have created in the past, specific data types, and metadata for existing datasets and code.
Finally, we could combine methods to develop new hybrid recommendation strategies.

A recent development in artificial intelligence is to build recommender systems using deep learning techniques (or \textit{deep recommenders})~\cite{Zhang2018, bar2020automatically}, particularly for content-based approaches.
Given our large available corpus of potential training data, unsupervised methods such as Recurrent Neural Networks~\cite{Goodfellow2016} could prove useful, since they are ideal for sequential data.
The \sys{} advisor model provides a useful framework from which to incorporate and merge future recommendation strategies for data science.
However, these topics are beyond the scope of this paper. 

\subsection{Limitations and Future Work}

Two participants from our formative study suggested that they would either appreciate being able to toggle off the recommendations or view them all without a filter. Based on these comments, it remains unclear how effective a code-free recommendation environment can be in teaching data science best practices. Thus, in our future work, we will test the strength of \sys{} recommendations and its effectiveness in teaching novice data scientists new techniques. 

Due to the many challenges of automatic code analysis, we currently do not allow users to write their own code directly in \sys{}, or even to modify existing code.
To make online code editing possible, we would need an automatic classification process that could determine how new code fits into the recommendation graph so that the system could resume the analysis with new recommendations after manual code \codeunit{}.
Such live updates to the recommender are not currently part of \sys{}, but are an interesting direction for future work.

We made several design decisions to the \sys{} notebook that will need to be revisited for a general implementation. 

\sys{} currently does not consider specifics about each input dataset while making recommendations---only display recommendations which do not programmatically fail to execute on the selected dataset.
This is a point of future work.

All of our analysis \codeunit{}s take a Pandas data frame as input, and generate a new data frame as output. 
Also, other disciplines use other data representations, and some computations may require passing multiple data objects as arguments. 
To address these limitations, we look to improving our existing design and thoroughly evaluating these improvements in our future work.

\bibliographystyle{abbrv-doi}

\bibliography{lodestar}
\end{document}